\documentclass[aps,prd,twocolumn,superscriptaddress,preprintnumbers,floatfix,nofootinbib,notitlepage,showkeys]{revtex4-1}

\usepackage[utf8]{inputenc}
\usepackage{cancel}

\usepackage{graphicx}
\usepackage{hyperref}
\usepackage{latexsym}
\usepackage{amsmath}
\usepackage{amssymb}
\usepackage{bbm}

\usepackage{ulem}
\usepackage{pdfsync}
\usepackage{epsfig}
\usepackage{epstopdf}
\usepackage{subfigure}
\usepackage{color}
\usepackage{comment}
\usepackage{slashed}

\def\beq{\begin{equation}}
\def\eeq{\end{equation}}
\def\baq{\begin{eqnarray}}
\def\eaq{\end{eqnarray}}

\newcommand{\be}{\begin{equation}} 
\newcommand{\ee}{\end{equation}}
\newcommand{\bea}{\begin{eqnarray}} 
\newcommand{\eea}{\end{eqnarray}}

\newcommand{\bmp}{\noindent\begin{minipage}{16cm}}
\newcommand{\emp}{\end{minipage}\vskip 7mm} 
\def\lsim{\mathrel{\raise.3ex\hbox{$<$\kern-.75em\lower1ex\hbox{$\sim$}}}}
\def\gsim{\mathrel{\raise.3ex\hbox{$>$\kern-.75em\lower1ex\hbox{$\sim$}}}}

\newcommand{\intron}[1]{}

\begin{document}

\title{Minimal Higgs inflation with an $R^2$ term in Palatini gravity}

\author{Tommi Tenkanen}
\email{ttenkan1@jhu.edu}
\affiliation{Department of Physics and Astronomy, Johns Hopkins University, \\
Baltimore, MD 21218, USA}

\begin{abstract}
It has recently been suggested that the Standard Model Higgs boson could act as the inflaton while {\it minimally} coupled to gravity -- given that the gravity sector is extended with an $\alpha R^2$ term and the underlying theory of gravity is of Palatini, rather than metric, type. In this paper, we revisit the idea and correct some shortcomings in earlier studies. We find that in this setup the Higgs can indeed act as the inflaton and that the tree-level predictions of the model for the spectral index and the tensor-to-scalar ratio are $n_s\simeq 0.941$, $r\simeq 0.3/(1+10^{-8}\alpha)$, respectively, for a typical number of e-folds, $N=50$, between horizon exit of the pivot scale $k=0.05\, {\rm Mpc}^{-1}$ and the end of inflation. Even though the tensor-to-scalar ratio is suppressed compared to the usual minimally coupled case and can be made compatible with data for large enough $\alpha$, the result for $n_s$ is in severe tension with the Planck results. We briefly discuss extensions of the model.
\end{abstract}

%
\maketitle

\section{Introduction}
\label{introduction}

The most minimal realization of inflation is the one where the Standard Model (SM) Higgs boson acts as the inflaton field, as such a scenario does not require any new fields on top of the ones we know to exist. The scenario was originally studied in Refs.~\cite{Salopek:1988qh,Bezrukov:2007ep} and it has gained significant attention during the last 10 years; for a recent review, see Ref. \cite{Rubio:2018ogq}. While alternative formulations exist \cite{Bauer:2008zj,Germani:2010gm,Nakayama:2010kt,Kamada:2010qe,Kamada:2012se,Jinno:2017lun,Enckell:2018kkc,Rasanen:2018ihz}, the usual and most studied scenario is based on the assumption that on top of its couplings to the other SM fields, the Higgs field is coupled {\it non-minimally} to gravity via $\xi \phi^2 R$, where $\phi$ is the Higgs boson, $R$ is the curvature scalar and $\xi$ is a dimensionless coupling constant. The coupling is required to sufficiently flatten the Higgs potential at large field values, so that slow-roll inflation and agreement with observations can be attained.

However, in choosing the form of such coupling, there is actually more freedom than is usually appreciated. In the usual metric formulation of gravity, one assumes that the space-time connection is determined uniquely by the metric only. In the so-called Palatini formalism, however, both the metric $g_{\mu\nu}$ and the connection $\Gamma$ are treated as independent variables, and the curvature scalar $R=g^{\mu\nu}R_{\mu\nu}(\Gamma)$ is actually a function of both $g_{\mu\nu}$ and $\Gamma$, as constructing the Ricci tensor $R_{\mu\nu}$ does not require the notion of a metric. One can take this approach also in the context of the General Relativity (GR), although in that case the constraint equation for the connection renders the two theories equivalent, i.e. in that case the metric and Palatini approaches provide for mere formulations of the same theory. However, with non-minimally coupled matter fields or otherwise enlarged gravity sector this is generally not the case~\cite{Sotiriou:2008rp}. In that case, one has to make a choice of the underlying gravitational degrees of freedom. Choosing the Palatini approach does not, however, constitute a modified theory of gravity any more than the metric one does, as currently we do not know what the underlying gravitational degrees of freedom are. Also, it does not necessarily amount to adding new degrees of freedom to the theory.  

In the context of inflation, this choice will generically change the dynamics and hence also the predictions of a given model when compared to alternative choices. This was originally noted in Ref. \cite{Bauer:2008zj} and has recently gained increasing attention \cite{Bauer:2010jg,Tamanini:2010uq,Rasanen:2017ivk,Tenkanen:2017jih,Racioppi:2017spw,Markkanen:2017tun,Jarv:2017azx,Racioppi:2018zoy,Enckell:2018kkc,Carrilho:2018ffi,Enckell:2018hmo,Antoniadis:2018ywb,Rasanen:2018fom,Kannike:2018zwn,Rasanen:2018ihz,Almeida:2018oid,Antoniadis:2018yfq,Takahashi:2018brt,Jinno:2018jei} (see also \cite{Azri:2017uor,Azri:2018gsz,Shimada:2018lnm}). Interestingly, it was recently pointed out in Ref. \cite{Antoniadis:2018yfq} that when the gravity sector is extended with an $\alpha R^2$ term (with $\alpha$ being a dimensionless parameter) and the underlying theory of gravity is of Palatini rather than metric type, the SM Higgs boson can act as the inflaton field while itself {\it minimally} coupled to gravity, i.e. even when the coupling $\xi\phi^2 R$ does not exist (for the case with a non-minimal coupling, see \cite{Enckell:2018hmo,Antoniadis:2018ywb}). This is a particularly interesting case, as a large non-minimal coupling to gravity renders the theory non-renormalizable at intermediate scales; see again Ref. \cite{Rubio:2018ogq}. Also, because both a non-minimal coupling of the type $\xi \phi^2 R$ and a Starobinsky-like $\alpha R^2$ term can be argued to be generated by quantum corrections in a curved background (see e.g. \cite{Salvio:2015kka}), also the inclusion of an $\alpha R^2$ term provides for a well-motivated starting point for the analysis of the physics at very high energies. It may indeed be that there is a sufficient hierarchy between the two operators, so that inflationary dynamics is dominated by the $\alpha R^2$ term rather than the non-minimal coupling $\xi \phi^2 R$, provided that the $\phi$ field indeed is the inflaton. 

In the following, we will study inflation in this setup. We will follow Ref. \cite{Antoniadis:2018yfq} in assuming that $\xi$ is small enough so that the non-minimal coupling does not play a role in determining the inflationary dynamics and the corresponding observables. We will, however, go beyond Ref. \cite{Antoniadis:2018yfq} in several different ways: we will compute the requirements for the model parameters so that they give the observed amplitude for the primordial curvature power spectrum, as well as discuss reheating and the allowed range for the number of e-folds between horizon exit of the pivot scale and the end of inflation. We will also correct some shortcomings in Ref. \cite{Antoniadis:2018yfq}.

The paper is organized as follows: in Section \ref{inflation}, we study inflation with an $\alpha R^2$ term in Palatini gravity and discuss the main observables. In Section \ref{reheating}, we discuss reheating and the number of e-folds relevant for the model under consideration, and present our main results. In Section \ref{conclusions}, we conclude.


\section{Inflation with an $R^2$ term}
\label{inflation}

We begin the discussion by presenting the model under consideration, following closely the discussion in Ref. \cite{Antoniadis:2018yfq}. We will study the following action:
\be \label{nonminimal_action1}
	S_J = \int d^4x \sqrt{-g}\left(\frac{1}{2}M_{\rm P}^2R + \frac{\alpha}{4} R^2 - \frac{1}{2} \nabla^{\mu}\phi\nabla_{\mu}\phi - V(\phi) \right) \,,
\ee
where $g$ is the determinant of the metric tensor $g_{\mu\nu}$, $\phi$ is the SM Higgs with the potential $V(\phi)$, $M_{\rm P}$ is the reduced Planck mass, $\alpha$ is a dimensionless parameter, $R=g^{\mu\nu}R_{\mu\nu}(\Gamma)$ is the Ricci scalar which depends on both the metric and the connection $\Gamma$ -- which we take to be an independent variable --, and $\nabla^\mu$ is the covariant derivative with respect to this connection\footnote{Note that the equation of motion for $\phi$ still involves the covariant derivative with respect to the Levi–Civita connection \cite{Rasanen:2017ivk}.}. We will assume that the possible non-minimal coupling between the Higgs and gravity is small enough not to take part in inflationary dynamics\footnote{For recent studies on mixed models where a scalar field couples non-minimally to an extended gravity sector, see \cite{Salvio:2015kka,Calmet:2016fsr,Wang:2017fuy,Ema:2017rqn,He:2018gyf,Ghilencea:2018rqg,Enckell:2018hmo,Antoniadis:2018ywb,Gundhi:2018wyz,Karam:2018mft,Antoniadis:2018yfq,Enckell:2018uic}.}.

For simplicity, we assume that the connection is torsion-free, $\Gamma^\lambda_{\alpha\beta}=\Gamma^\lambda_{\beta\alpha}$ (for non-vanishing torsion, see \cite{Rasanen:2018ihz,Shimada:2018lnm}). As discussed in Section \ref{introduction}, in the context of GR the constraint equation for the connection imposes $\Gamma$ to be the Levi-Civita connection and hence renders the two formalisms -- metric and Palatini --  equivalent. However, with non-minimally coupled matter fields or otherwise enlarged gravity sector this is generally not the case~\cite{Sotiriou:2008rp}, and as we currently do not know what the fundamental gravitational degrees of freedom are, it is a natural starting point to consider a theory where the connection (and hence space-time geometry) is determined by both the metric and the matter fields. As we will see, this has very interesting consequences.

The action \eqref{nonminimal_action1} can be written dynamically equivalently as~\cite{Sotiriou:2008rp}
\bea
	S_J &=& \int d^4x \sqrt{-g}\bigg (\frac{1}{2}M_{\rm P}^2\left(1+\alpha z^2 \right) R -\frac{\alpha}{4}z^4 \\ \nonumber
	&-& \frac{1}{2}\nabla^{\mu}\phi\nabla_{\mu}\phi - V(\phi) \bigg )\,,
\eea
where $z$ is an auxiliary field. Performing a Weyl transformation
\be \label{Omega1}
	g_{\mu\nu} \to \Omega g_{\mu\nu}, \hspace{1cm} \Omega \equiv 1+\frac{\alpha z^2}{M_{\rm P}^2} \,,
\ee
gives the Einstein frame action
\be \label{einsteinframe1}
	S_E = \int d^4x \sqrt{-g}\left(\frac{1}{2}M_{\rm P}^2 R - \frac{1}{2\Omega}\nabla^{\mu}\phi\nabla_{\mu}\phi - \frac{V(\phi)}{\Omega^2} \right) \,,
\ee
with a canonical gravity sector. Because now the connection appears only in the usual Einstein-Hilbert term, in this frame we retain the Levi-Civita connection, which shows that the model \eqref{nonminimal_action1} actually belongs to a class of {\it metric-affine} theories \cite{Azri:2017uor,Azri:2018gsz,Rasanen:2018ihz,Shimada:2018lnm}. Varying the action \eqref{einsteinframe1} with respect to the field $z$ gives an algebraic constraint equation -- instead of an equation of motion containing derivatives of $z$, as in the Palatini case the field is non-dynamical --, which is solved for \cite{Enckell:2018hmo,Antoniadis:2018ywb,Antoniadis:2018yfq}
\be
\frac{z^2}{M_{\rm P}^2} = \frac{4V(\phi) + \nabla^{\mu}\phi\nabla_{\mu}\phi}{M_{\rm P}^4 - \alpha \nabla^{\mu}\phi\nabla_{\mu}\phi} .
\ee
Substituting this into Eq. \eqref{einsteinframe1} gives
\be \label{einsteinframe2}
	S_E \simeq \int d^4x \sqrt{-g}\left(\frac{1}{2}M_{\rm P}^2 R - \frac{\nabla^{\mu}\phi\nabla_{\mu}\phi}{1 + 4\alpha \frac{V(\phi)}{M_{\rm P}^4}} - \frac{V(\phi)}{1 + 4\alpha \frac{V(\phi)}{M_{\rm P}^4}} \right) \,,
\ee
where we assumed $\dot{\phi}^2 \ll V(\phi)$, i.e. the field is in slow-roll and that spatial gradients of the field vanish everywhere during inflation.

We now assume that the Higgs potential during inflation is $V(\phi) = \lambda \phi^4/4$, which holds for large field values, $\phi \gg v$, where $v=246$ GeV is the Higgs' vacuum expectation value at zero temperature. With a suitable field redefinition $\phi = \phi(\chi)$, determined by 
\be \label{chi1}
	\frac{{\rm d}\phi}{{\rm d}\chi} = \sqrt{1+ \lambda\alpha \left(\frac{\phi}{M_{\rm P}}\right)^4} \,,
	\ee
the kinetic term for the Higgs can be written in a canonical form. The solution to Eq. \eqref{chi1} can be written in terms of the elliptic integral of the first kind but the result is not particularly enlightening. The action~\eqref{einsteinframe2} then becomes
\be \label{EframeS1}
	S_{\rm E} = \int d^4x \sqrt{-g}\bigg(\frac{1}{2}M_{\rm P}^2R -\frac{1}{2}{\partial}_{\mu}\chi{\partial}^{\mu}\chi - U(\chi)  \bigg) \,,
\ee
where 
\be
\label{Uchi}
U(\chi) = \frac{\lambda \phi^4(\chi)}{4\left(1+\lambda\alpha\frac{\phi^4(\chi)}{M_{\rm P}^4} \right)} ,
\ee 
where $\phi(\chi)$ is given by Eq. \eqref{chi1}. The result \eqref{Uchi} is the vanishing coupling limit of the more general result presented in Refs. \cite{Enckell:2018hmo,Antoniadis:2018ywb}, where a non-minimal coupling between the inflaton field and gravity was assumed. Note that when $\chi\to 0$, the usual Einstein-Hilbert gravity, i.e. pure GR, is retained and the assumption of the underlying theory of gravity being of Palatini type loses its speciality; it is, in this limit, simply an alternative way to formulate GR  -- albeit equally suitable as the usual metric formulation, as discussed above. 

\begin{figure}
\begin{center}
\includegraphics[width=.485\textwidth]{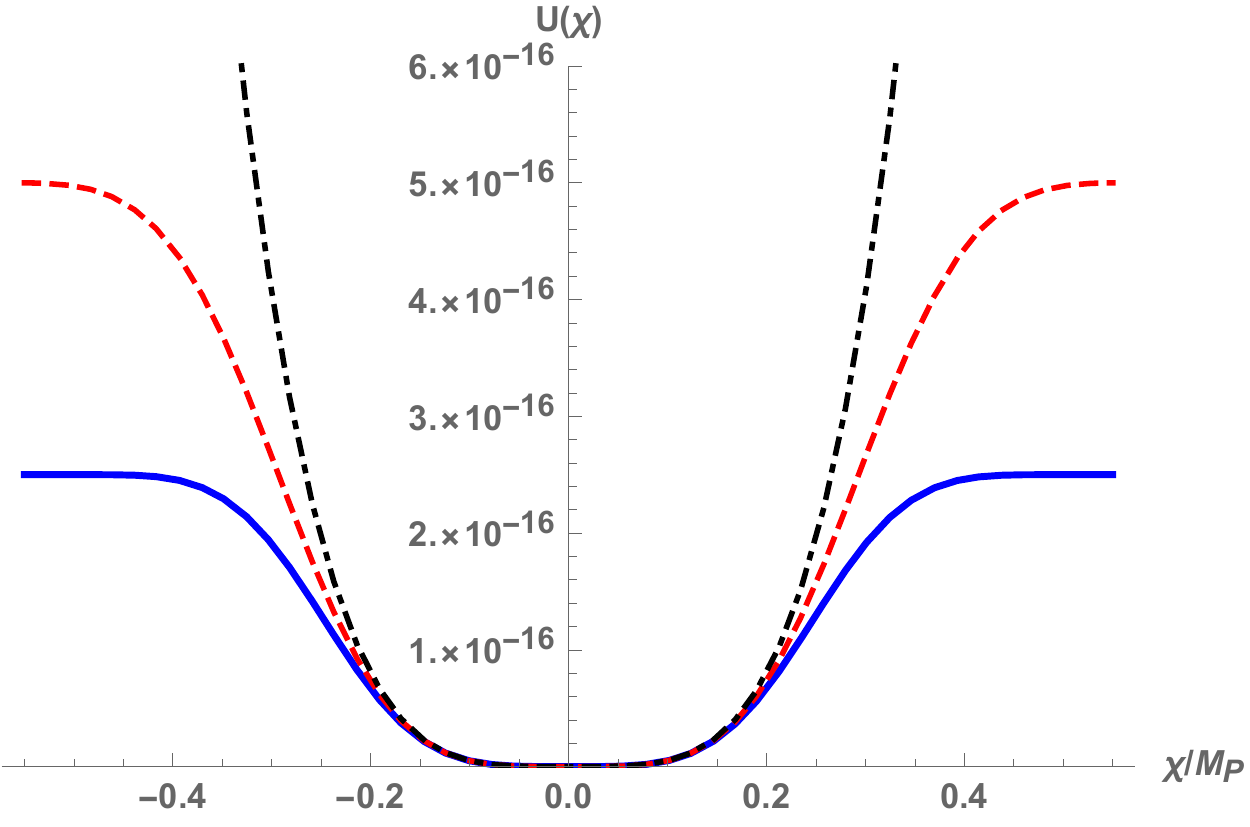}
\includegraphics[width=.485\textwidth]{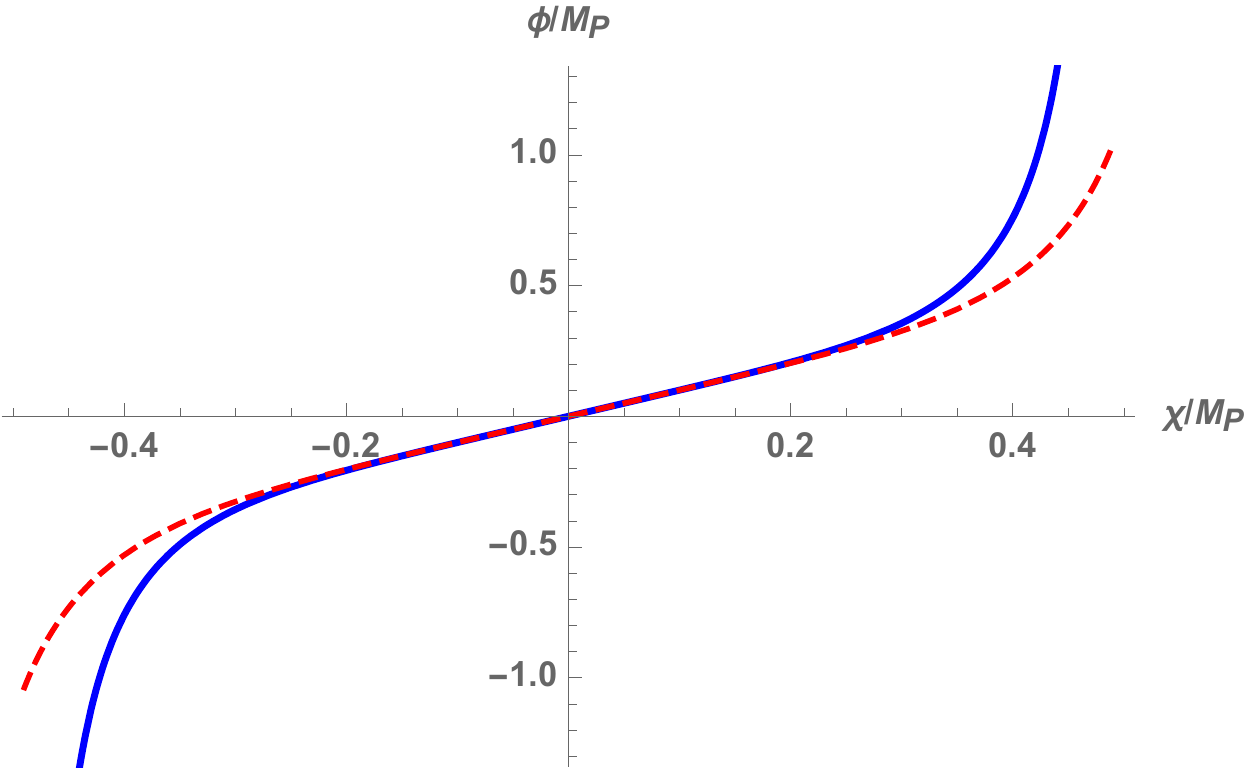}
\caption{{\it Upper panel}: The Einstein frame potential for varying $\alpha$. {\it Lower panel}: The Jordan frame field $\phi$ vs. the canonicalized Einstein frame field $\chi$. In this Figure $\alpha=10^{15}$ (blue thick curves) and $\alpha=5\times 10^{14}$ (red dashed curves). The quartic self-coupling is $\lambda=2\times 10^{-13}$ for all curves. Note that the field reaches a plateau at values smaller than the Planck scale, $\chi < M_{\rm P}$. The black dot-dashed curve in the middle of the upper panel is the usual quartic potential, $\lambda\chi^4/4$, shown here for reference.
}
\label{fig:potential_plots}
\end{center}
\end{figure}

In the upper panel of Fig.~\ref{fig:potential_plots}, we show how the Einstein frame potential depends on the value of $\alpha$ for a fixed $\lambda$. We see that for suitable choices of parameters, the potential develops a plateau at large field values and hence exhibits behavior which is very suitable for slow-roll inflation. Notably, the plateau can be reached at values smaller than the Planck scale, $\chi < M_{\rm P}$. However, calculating the observables is non-trivial, as also the rate of change of the field is modified, see Eq. \eqref{chi1}. In the lower panel of Fig.~\ref{fig:potential_plots}, we show the mapping between the Jordan frame field $\phi$ and the canonicalized Einstein frame field $\chi$.

As shown in Ref. \cite{Enckell:2018hmo}, the inflationary dynamics is characterized by the usual slow-roll parameters
\begin{eqnarray}
\label{SRparameters1}
\epsilon &\equiv& \frac12 M_{\rm P}^2 \left(\frac{U'}{U}\right)^2 =  \frac{1}{1+\alpha\lambda\left(\frac{\bar{\phi}}{M_{\rm P}}\right)^4}{\bar{\epsilon}} \,, \nonumber \\
\eta &\equiv& M_{\rm P}^2 \frac{U''}{U} = \bar{\eta}-3\frac{\alpha\lambda\left(\frac{\bar{\phi}}{M_{\rm P}}\right)^4}{1+\alpha\lambda\left(\frac{\bar{\phi}}{M_{\rm P}}\right)^4}\bar{\epsilon} \,, 
\end{eqnarray}
where the prime denotes derivative with respect to $\chi$ and $\bar{\epsilon}, \bar{\eta}$ are the slow-roll parameters and $\bar{\phi}$ the field value in the case of $\alpha=0$. Because everything here has been expressed in terms of quantities of the well-known case of $V=\lambda\bar{\phi}^4/4$ inflation (in the following, '$\lambda\phi^4$ inflation' for short), we can use the canonical results of this model to find the the leading order expressions for the main inflationary observables, i.e. the spectral index and the tensor-to-scalar ratio. The number of $e$-folds between the horizon exit of a given scale and the end of inflation, $N$, is in that case given by
\be \label{Ndef}
	N = \int_{\bar{\phi}_f}^{\bar{\phi}_i} \frac{{\rm d}\bar{\phi}}{M_{\rm P}} \, \frac{1}{\sqrt{2\bar{\epsilon}}}\,,
\ee
where $\bar{\phi}_i$ is the field value when the given scale exited the horizon. The field value at the end of inflation, $\bar{\phi}_f$, is defined by $\bar{\epsilon}(\bar{\phi}_f)=1$. For $V=\lambda\bar{\phi}^4/4$, we thus find in the standard manner $\bar{\epsilon}= 8M_{\rm P}^2/\bar{\phi}_i^2 = 1/(N+1)$ and $\bar{\eta} =3\bar{\epsilon}/2$. The leading order expressions for the spectral index and the tensor-to-scalar ratio therefore are
\begin{eqnarray}
\label{nsralpha}
n_s &=& 1 - 6\epsilon + 2\eta = 1  - \frac{3}{N+1}\,, \\ \nonumber
r &=& 16\epsilon = \frac{16}{\left(N+1\right)\left(1+64\alpha\lambda(N+1)^2\right)} \,,
\end{eqnarray}
respectively. We see that the expression for $n_s$ is in this case exactly the same as the one in the usual $\lambda\phi^4$ inflation. Indeed, as discussed in Ref. \cite{Enckell:2018hmo}, all quantities defined as derivatives of the power spectrum, such as the spectral index, its running and the running of the running, remain unaffected to leading order, because the curvature power spectrum remains the same. Its amplitude is given by ~\cite{Lyth:1998xn,Aghanim:2018eyx}
\be
\label{cobe}
\mathcal{P}_{\zeta} = \frac{1}{24 \pi^2 M_{\rm P}^4} \frac{U}{\epsilon} = \frac{2(N+1)^3\lambda}{3 \pi^2},
\ee
with the observed value $\mathcal{P}_{\zeta}=2.1\times 10^{-9}$ \cite{Akrami:2018odb}. Furthermore, the most recent analysis of observations of the CMB made by the Planck satellite also give (at the $68\%$ confidence level) \cite{Akrami:2018odb}
\be
n_s = 0.9625\pm 0.0048, 
\label{Planck_const_run}
\ee
for the TT,TE,EE+lowE+lensing dataset. The joint analysis of data by Planck and BICEP2/Keck Array gives (at the $95\%$ confidence level) \cite{Ade:2018gkx}
\be
\label{planck_r}
r<0.06 .
\ee  
The observables are constrained at the pivot scale $k_*=0.05\, {\rm Mpc}^{-1}$. We see from Eq. \eqref{cobe} that at tree-level, the value of $\alpha$ does not enter the expression for $\mathcal{P}_{\zeta}$ at all and the result indeed remains unchanged from the one of $\lambda\phi^4$ inflation. One can see that for $N\gtrsim 50$, the requirement for the quartic self-coupling is $\lambda \lesssim \mathcal{O}(10^{-12})$, which in the case of the SM Higgs requires considerable fine-tuning of the SM $\beta$-functions. This is not entirely inconceivable (see e.g. \cite{Degrassi:2012ry}) but may decrease the appeal of the most minimal model. However, we will see that this is not the only problematic thing in this model.

Before discussing the number of e-folds in more detail, let us note that unlike the usual $\lambda\phi^4/4$ inflation, the present model delivers a prediction for $r$ which can be made compatible with the Planck data for large enough $\alpha$. Solving for $\lambda$ from Eq. \eqref{cobe} and substituting it into Eq. \eqref{nsralpha}, we find
\be
r = \frac{16}{N+1+96\pi^2\mathcal{P}_{\zeta}\alpha} .
\ee
We see that for e.g. $N=50$, the value of $\alpha$ which makes the prediction compatible with the Planck results is in the $\mathcal{O}(10^{8})$ ballpark. This should not be seen as a hindrance, though, as $\alpha$ can be seen as a manifestation of a new scale $M\equiv M_{\rm P}/\alpha^{1/4}<M_{\rm P}$, where the high energy physics start exhibiting phenomena different from those at low energies, rather than as a coupling constant which should be small. In Ref. \cite{Antoniadis:2018yfq} it was assumed that $\sqrt{\lambda\alpha}\leq 10^{-2}$ always but if $\alpha<1$, the new scale is $M>M_{\rm P}$, which is possibly in conflict with quantum gravity. In this paper, we will therefore assume $\alpha > 1$.


\section{Reheating and the number of e-folds}
\label{reheating}

To find accurate predictions for the main inflationary observables discussed above, we will now discuss reheating and the number of e-folds between horizon exit of the pivot scale and the end of inflation. The latter is required for solving the field value at which the perturbations we observe in the CMB are generated, as given by Eq. \eqref{Ndef}.

The number of e-folds between horizon exit of the pivot scale and the end of inflation is
\be
N = \ln\left(\frac{a_{\rm end}}{a_{\rm reh}}\right) + \ln\left(\frac{a_{\rm reh}}{a_0}\right) + \ln\left(H_k k^{-1}\right) ,
\ee
where $a$ refers to the scale factor at different times (end of inflation, reheating, and today, respectively) and $H_k$ is the Hubble parameter during inflation at the time the scale $k$ exited the horizon. In the following, we will be interested in the Planck pivot scale $k=k_*=0.05\, {\rm Mpc}^{-1}$. Assuming a constant equation of state during reheating, the ratio of the scale factors at the end of inflation and at the time of reheating is
\bea
&\ln& \left(\frac{a_{\rm end}}{a_{\rm reh}}\right) = \ln\left(\left(\frac{\rho_{\rm reh}}{\rho_{\rm end}}\right)^{\frac{1}{3(1+w)}}\right)  \\ \nonumber
&=& \frac{1}{3(1+w)}\left(\ln\left(\frac{\pi^2 g_*(T_{\rm reh})}{90c^2}\right) + 4\ln\left(\frac{T_{\rm reh}}{M_{\rm P}}\right)
 - 2\ln\left(\frac{H_k}{M_{\rm P}}  \right) \right),
\eea
where $\rho$ is the total energy density at a given time, $w$ is the average equation of state parameter during reheating, $c\equiv H_{\rm end}/H_k$ characterizes how much $H$ changes between horizon exit of the scale $k$ and the end of inflation, $T_{\rm reh}$ is the reheating temperature, and $g_*$ is the effective number of degrees of freedom at the time of reheating. The maximum reheating temperature is
\bea
\label{TrehMax}
T_{\rm reh}^{\rm max} &=& \left(\frac{90}{\pi^2 g_*(T_{\rm reh})}\right)^{1/4}\sqrt{H_{\rm end}M_{\rm P}} \\ \nonumber
&\simeq& 7.5\times 10^{15}\left(\frac{r}{0.1}\right)^{1/4}\,{\rm GeV} ,
\eea
which assumes instant reheating after inflation\footnote{The absolute minimum reheating temperature is given by Big Bang Nucleosynthesis, $T\sim 1$ MeV.}. We also have
\be
\ln\left(\frac{a_{\rm reh}}{a_0}\right) = \frac13\ln\left(\frac{g_*(T_0)T_0^3}{g_*(T_{\rm reh})T^3_{\rm reh}} \right) \simeq -67 - \ln\left(\frac{T_{\rm reh}}{10^{16}\,{\rm GeV}}\right),
\ee
where $T_0=2.725$ K is the CMB temperature today, and
\be
\ln\left(H_k k^{-1}\right) \simeq 123 + \frac12\ln\left(\frac{r}{0.1}\right) - \ln\left(\frac{k}{0.05\,{\rm Mpc}^{-1}}\right) ,
\ee
where we used
\be
H_k = \sqrt{\frac{\pi r P_\zeta}{2}}M_{\rm P} ,
\ee
which follows from the definition of the tensor-to-scalar ratio $r\equiv P_T/P_\zeta$, where $P_T = 8(H_k/2\pi)^2M_{\rm P}$. The overall result is not particularly sensitive\footnote{For example, even $g_*(T_{\rm reh})=10^5$ does not change the result by more than $\Delta N\sim 7$. Also, because $|\dot{H}|=\epsilon H^2$ during slow-roll, the Hubble scale is not expected to change much during the last stages of inflation, i.e. between horizon exit of the observable scales and the end of inflation.} to changes in $c$ and $g_*(T_{\rm reh})$, so we can take $g_*(T_{\rm reh})\sim 100$ and $c\sim 1$. Assuming that $w\geq -1/3$ during reheating (which is a very plausible assumption, as typically $w\geq 0$ \cite{Podolsky:2005bw,Munoz:2014eqa}), we find that the maximum number of e-folds between horizon exit of the pivot scale $k_*=0.05\, {\rm Mpc}^{-1}$ and the end of inflation is
\be
N^{\rm max}_{0.05} \simeq 56 + \ln\left(\frac{T_{\rm reh}}{10^{16}\,{\rm GeV}}\right) .
\ee
For the other pivot scale used by Planck, $k=0.002\, {\rm Mpc}^{-1}$, the first term is 59. The result is in line with e.g. Refs. \cite{Dodelson:2003vq,Liddle:2003as,Munoz:2014eqa}, and shows that the results presented in the original paper on Palatini-$R^2$ inflation with a Higgs-like scalar field \cite{Antoniadis:2018yfq}, where $N\geq 65$ was assumed, require implausible assumptions of the thermal history of the Universe after inflation in this model. Note that this conclusion is independent of reheating mechanism.

Using the above results, we can find a plausible range for the number of e-folds in the model under consideration. Assuming the SM particle content, $g_*(T_{\rm reh})=106.75$, and taking $w\simeq 0$, $c=1$, we obtain 
\be
N_{0.05} \simeq 56 + \frac13\ln\left(\frac{T_{\rm reh}}{10^{16}\,{\rm GeV}}\right) + \frac16\ln\left(\frac{r}{0.1}\right).
\ee
By varying $T_{\rm reh}$ between $10^6$ GeV and the maximum reheating temperature \eqref{TrehMax}, we find $N\simeq 45...56$. This gives relatively accurate predictions for the main inflationary observables, as discussed above. The result is not particularly sensitive to the values of $w$ or $c$.

\begin{table}
\begin{center}
\normalsize
\begin{tabular}{| c | c | c | c | c |}
\hline
& $n_s$  & r ($\alpha = 10^8$) & r ($\alpha = 10^{12}$) & r ($\alpha = 10^{16}$)  \\
\hline
$N=56$ & $0.947$ & $0.06$ & $8\times 10^{-6}$ & $8\times 10^{-10}$  \\ 
\hline
$N=45$ & $0.935$ & $0.07$ & $8\times 10^{-6}$  & $8\times 10^{-10}$ \\ 
\hline
\end{tabular}
\caption{Predictions for the spectral index and the tensor-to-scalar ratio for the given number of e-folds. The values of $N$ shown here represent the upper and lower limits discussed in Section \ref{reheating}.}
\label{table:observables}
\end{center}
\end{table}

The results are shown in Table \ref{table:observables}. We see that for the given range of e-folds, predictions of the model are in severe tension with the Planck results, Eq. \eqref{Planck_const_run}. While the predicted value for the tensor-to-scalar ratio can be made compatible with the Planck data for large enough $\alpha$, making the scenario compatible with the limits on $n_s$ requires implausible assumptions of the reheating period. However, the above results apply only at tree-level. Because in the present scenario the Higgs is coupled only minimally to gravity, it would be interesting to see if the SM $\beta$-functions could be applied to the present scenario as such, and if so, to what extent they can change the tree-level analysis conducted in this study.


\section{Conclusions}
\label{conclusions}

We have studied a scenario where the SM Higgs boson acts as the inflaton field while minimally coupled to gravity. This is possible when the gravity sector is extended with an $R^2$ term and the underlying theory of gravity is of Palatini type. We corrected some shortcomings in previous studies and computed the predictions of the model for the spectral index and the tensor-to-scalar ratio. We found $n_s\simeq 0.941$, $r\simeq 0.3/(1+10^{-8}\alpha)$, respectively, for a typical number of e-folds, $N=50$, between horizon exit of the pivot scale $k=0.05\, {\rm Mpc}^{-1}$ and the end of inflation. While the predicted value for the tensor-to-scalar ratio can be made compatible with the Planck data for large enough $\alpha$, making the scenario compatible with the limits on $n_s$ requires implausible assumptions of the reheating period.  However, going beyond the tree-level analysis conducted in this paper may change the predictions of the model without further modifications of the underlying theory.


\section*{Acknowledgements}
I thank Marc Kamionkowski, Syksy R\"as\"anen, Eemeli Tomberg, and Ville Vaskonen for discussions and valuable comments on the manuscript, as well as Guillem Dom\`{e}nech, Jos\'{e} Ram\'{o}n Espinosa, Lingyuan Ji, Alexandros Karam, and Angelos Lykkas for correspondence and discussions. I thank the University of Helsinki for hospitality and acknowledge the Simons foundation for funding.


\bibliography{minimal_HI}


\end{document}